\documentclass
[preprint,showpacs,a4paper,pra,amsfonts,onecolumn,noshowkeys]{revtex4}%
\usepackage{amsfonts}
\usepackage{amsmath}
\usepackage{amssymb}
\usepackage{graphicx}%
\begin{document}
\title{Simple scheme for two-qubit Grover search in cavity QED}
\author{Z.J. Deng$^{1,2}$}
\email{dengzhijiao926@hotmail.com}
\author{M. Feng$^{1}$}
\author{K.L. Gao$^{1}$}
\affiliation{$^{1}$State Key Laboratory of Magnetic Resonance and Atomic and Molecular
Physics, Wuhan Institute of Physics and Mathematics, Chinese Academy of
Sciences, Wuhan 430071, China}
\affiliation{Centre for Cold Atom Physics, Chinese Academy of Sciences, Wuhan 430071, China}
\affiliation{$^{2}$Graduate School of the Chinese Academy of Sciences, Beijing 100049, China}

\begin{abstract}
Following the proposal by F. Yamaguchi et al.[Phys. Rev. A \textbf{66}, 010302
(R) (2002)], we present an alternative way to implement the two-qubit Grover
search algorithm in cavity QED. Compared with F. Yamaguchi et al.'s proposal,
with a strong resonant classical field added, our method is insensitive to
both the cavity decay and thermal field, and doesn't require that the cavity
remain in the vacuum state throughout the procedure. Moreover, the qubit
definitions are the same for both atoms, which makes the experiment easier.
The strictly numerical simulation shows that our proposal is good enough to
demonstrate a two-qubit Grover's search with high fidelity.

\end{abstract}

\maketitle

The Grover search algorithm\cite{1} is an efficient quantum algorithm to look
for one item in an unsorted datebase of size $N$. While the most efficient
classical algorithm which examines items one by one needs on average $N/2$
queries, the Grover's quantum algorithm uses only $O(\sqrt{N})$ queries to
accomplish the same task. The efficiency of this algorithm has been tested
experimentally in few-qubit cases by NMR \cite{2} and by optics \cite{3}.

Grover's search can be briefly described as follows: For items represented by
the computational states $|X\rangle$ with $X=0,1...N-1,$ in a quantum register
with $n$ qubits, we have $N=2^{n}$ possible states. The search starts from a
superposition state $|\Psi_{0}\rangle=\frac{1}{\sqrt{N}}\overset
{N-1}{\underset{X=0}{\sum}}|X\rangle$, where each item has an equal
probability to get picked. One search step (i.e., a query) includes two key
operations \cite{4}: (i) Inverting the amplitute of the target item; (ii)
Performing a diffusion transform $D$, i.e., inversion about the average state
$|\Psi_{0}\rangle$ with $D_{ij}=2/N$ for $i\neq j$ and $D_{ii}=-1+2/N$ . If
the target item is $|\tau\rangle,$ then the operation in case (i) results in a
conditional phase gate $I_{\tau}=$ $I-2|\tau\rangle\langle\tau|,$ where $I$ is
the $N\times N$ identity matrix. After O($\sqrt{N}$) queries, the amplitude of
the identified target would be amplified while amplitude of non-target items
are shrunk to be negligible. Thus we get the target item with high probability.

Various schemes have been proposed for implementing several quantum algorithms
in cavity QED, for example, Grover search algorithm\cite{5}, quantum discrete
Fourier transform\cite{6}, Deutsch-Jozsa algorithm\cite{7}, quantum dense
coding\cite{8} and so on. Although cavity QED is one of the qualified
candidates for quantum information processing(QIP) and attracts much
attention, decoherence of the cavity field remains to be a big obstacle for
QIP in cavity QED. Recently, Zheng and Guo proposed two atoms interacting with
a nonresonant cavity, in which the two atoms can be entangled without
information transfer between atoms and the cavity\cite{9}. However, it
requires the cavity to be initially in the vacuum state. Osnaghi et
al.\cite{10} have experimentally demonstrated this scheme. Yamaguchi et al.
extended Zheng and Guo's proposal to a scheme to realize the two-qubit Grover
search algorithm in cavity QED. In order to achieve the quantum phase gate,
they have to choose different levels for qubit encoding for the two
atoms\cite{5}. In Refs. \cite{8,11}, with a strong resonant classical field
added, the photon-number-dependent Stark shift can be canceled. It does not
require that the cavity be initially in the vacuum state, and the scheme is
insensitive to both the cavity decay and the thermal field.

In this paper, we modify the proposal in Ref. \cite{5} by adding a strong
resonant classical field. Comparing with Ref. \cite{5}, we find our method
have following merits: (i) Initial vacuum cavity field is not needed, and our
method is insensitive to the thermal field besides cavity decay; (ii) Qubit
definitions are the same for the two atoms, which makes the experiment easier;
(iii) The two-qubit gate can be easily achieved by an appropriate Rabi
frequency $\Omega$ (defined below); (iiii) Except for two NOT gates on atom 2
for labeling target state $|g_{1}\rangle|e_{2}\rangle$ or $|e_{1}\rangle
|g_{2}\rangle$, all the other operations are simultaneously imposed on the two
atoms, which makes the implementation more compact.

We consider two identical two-level atoms simultaneously interacting with a
single-mode cavity field and driven by a classical field. The Hamiltonian
(assuming $\hbar=1$) in the rotating-wave approximation reads\cite{8,11}%

\begin{equation}
H=\frac{1}{2}\overset{2}{\underset{j=1}{\sum}}\omega_{0}\sigma_{z,j}%
+\omega_{a}a^{+}a+\overset{2}{\underset{j=1}{\sum}}[g(a^{+}\sigma_{j}%
^{-}+a\sigma_{j}^{+})+\Omega(\sigma_{j}^{+}e^{-i\omega t}+\sigma_{j}%
^{-}e^{i\omega t})] \label{1}%
\end{equation}

where $\sigma_{z,j}=|e_{j}\rangle\langle e_{j}|-|g_{j}\rangle\langle g_{j}|$,
$\sigma_{j}^{+}=|e_{j}\rangle\langle g_{j}|$, $\sigma_{j}^{-}=|g_{j}%
\rangle\langle e_{j}|$, with $|e_{j}\rangle$ ( $|g_{j}\rangle$ ) being the
excited ( ground ) state of the $j$th atom. $\omega_{0}$, $\omega_{a}$,
$\omega$ are the frequency for atomic transition, cavity mode and classical
field respectively. $a^{+}$, $a$ are the creation and annihilation operators
for the cavity mode. $g$ is the atom-cavity coupling strength and $\Omega$ is
the Rabi frequency of the classical field. Assuming $\omega_{0}=\omega$ and
$\delta=\omega_{0}-\omega_{a},$ we have following Hamiltonian in the
interaction picture \cite{8,11}%

\begin{equation}
H_{I}=\overset{2}{\underset{j=1}{\sum}}[\Omega(\sigma_{j}^{+}+\sigma_{j}%
^{-})+g(e^{-i\delta t}a^{+}\sigma_{j}^{-}+e^{i\delta t}a\sigma_{j}^{+})]
\label{2}%
\end{equation}

When $\Omega\gg$ $\delta,$ $g$ and $\delta\gg g$, we can get the evolution
operator of the system in the interaction picture\cite{8,11}%

\begin{equation}
U_{I}(t)=e^{-iH_{0}t}e^{-iH_{e}t} \label{3}%
\end{equation}

\bigskip\ with%
\begin{equation}
H_{0}=\overset{2}{\underset{j=1}{\sum}}\Omega(\sigma_{j}^{+}+\sigma_{j}^{-})
\label{4}%
\end{equation}

\bigskip%
\begin{equation}
H_{e}=\lambda\lbrack\frac{1}{2}\overset{2}{\underset{j=1}{\sum}}(|e_{j}%
\rangle\langle e_{j}|+|g_{j}\rangle\langle g_{j}|)+(\sigma_{1}^{+}\sigma
_{2}^{+}+\sigma_{1}^{+}\sigma_{2}^{-}+H.c.)] \label{5}%
\end{equation}

where $\lambda=g^{2}/2\delta.$ It's obvious from Eq. (5), the phonon-number
dependent Stark shift has been canceled by an additional strong resonant
classical field. If we define $J_{x}=\frac{1}{2}\overset{2}{\underset
{j=1}{\sum}}(\sigma_{j}^{+}+\sigma_{j}^{-}),$ Eqs. (4) and (5) reduce to
$H_{0}=2\Omega J_{x}$, $H_{e}=2\lambda J_{x}^{2}$ respectively and Eq. (3) becomes%

\begin{align}
U_{I}(t)  &  =e^{-i2\Omega tJ_{x}}e^{-i2\lambda tJ_{x}^{2}}\label{6}\\
&  =e^{-ib(J_{x}^{2}+hJ_{x})}\text{ \ \ \ \ }\nonumber
\end{align}

with $b=2\lambda t$ $\ $and$\ h=\frac{\Omega}{\lambda}$. In the subspace
spanned by $|e_{1}\rangle|e_{2}\rangle,$ $|e_{1}\rangle|g_{2}\rangle,$
$|g_{1}\rangle|e_{2}\rangle,$ $|g_{1}\rangle|g_{2}\rangle,$ we define the
two-qubit Hadamard gate%

\begin{align}
H^{\otimes2}  &  =\overset{2}{\underset{i=1}{\Pi}}H_{i}\label{7}\\
&  =(\frac{1}{\sqrt{2}})^{2}\left[
\begin{array}
[c]{cc}%
-1 & 1\\
1 & 1
\end{array}
\right]  \otimes\left[
\begin{array}
[c]{cc}%
-1 & 1\\
1 & 1
\end{array}
\right] \nonumber\\
&  =\frac{1}{2}\left[
\begin{array}
[c]{cccc}%
1 & -1 & -1 & 1\\
-1 & -1 & 1 & 1\\
-1 & 1 & -1 & 1\\
1 & 1 & 1 & 1
\end{array}
\right] \nonumber
\end{align}

where $H_{i}$ is the Hadamard gate acting on the ith atom, transforming states
as $|g_{i}\rangle\rightarrow\frac{1}{\sqrt{2}}(|g_{i}\rangle+|e_{i}\rangle)$,
$|e_{i}\rangle\rightarrow\frac{1}{\sqrt{2}}(|g_{i}\rangle-|e_{i}\rangle)$. The
$U_{I}(t)$ can be expressed in the same basis as:%

\begin{equation}
U_{I}(t)=\left[
\begin{array}
[c]{cccc}%
\frac{1}{2}+\frac{1}{2}\cos(bh)e^{-ib} & \frac{-i}{2}\sin(bh)e^{-ib} &
\frac{-i}{2}\sin(bh)e^{-ib} & \frac{-1}{2}+\frac{1}{2}\cos(bh)e^{-ib}\\
\frac{-i}{2}\sin(bh)e^{-ib} & \frac{1}{2}+\frac{1}{2}\cos(bh)e^{-ib} &
\frac{-1}{2}+\frac{1}{2}\cos(bh)e^{-ib} & \frac{-i}{2}\sin(bh)e^{-ib}\\
\frac{-i}{2}\sin(bh)e^{-ib} & \frac{-1}{2}+\frac{1}{2}\cos(bh)e^{-ib} &
\frac{1}{2}+\frac{1}{2}\cos(bh)e^{-ib} & \frac{-i}{2}\sin(bh)e^{-ib}\\
\frac{-1}{2}+\frac{1}{2}\cos(bh)e^{-ib} & \frac{-i}{2}\sin(bh)e^{-ib} &
\frac{-i}{2}\sin(bh)e^{-ib} & \frac{1}{2}+\frac{1}{2}\cos(bh)e^{-ib}%
\end{array}
\right]  \label{8}%
\end{equation}

If we choose $b=\frac{\pi}{2}$, $bh=\frac{\pi}{2}+2m\pi$ ($m$ is an interger),
i.e., $\lambda t=\frac{\pi}{4},\frac{\Omega}{\lambda}=4m+1,$ we can get%

\begin{equation}
U_{I}(t_{D})=\left[
\begin{array}
[c]{cccc}%
\frac{1}{2} & \frac{-1}{2} & \frac{-1}{2} & \frac{-1}{2}\\
\frac{-1}{2} & \frac{1}{2} & \frac{-1}{2} & \frac{-1}{2}\\
\frac{-1}{2} & \frac{-1}{2} & \frac{1}{2} & \frac{-1}{2}\\
\frac{-1}{2} & \frac{-1}{2} & \frac{-1}{2} & \frac{1}{2}%
\end{array}
\right]  =-D \label{9}%
\end{equation}

where $t_{D}=\frac{\pi}{4\lambda}$\cite{12}. So by choosing an appropriate
value of $\Omega,$ we can generate a two-qubit diffusion transform $D$
(different by $-1$ prefactor). The two-qubit conditional phase gate to label
different target states will also be generated in a natrural way. It's easy to have%

\begin{equation}
H^{\otimes2}U_{I}(t)H^{\otimes2}=\left[
\begin{array}
[c]{cccc}%
e^{ib(h-1)} & 0 & 0 & 0\\
0 & 1 & 0 & 0\\
0 & 0 & 1 & 0\\
0 & 0 & 0 & e^{-ib(h+1)}%
\end{array}
\right]  \label{10}%
\end{equation}

If we choose $b=\frac{\pi}{2}$, $h=4m+1$ ($m$ is an interger), i.e.,
$t_{1}=\frac{\pi}{4\lambda},\frac{\Omega}{\lambda}=4m+1,$ we have%

\begin{equation}
H^{\otimes2}U_{I}(t_{1})H^{\otimes2}=\left[
\begin{array}
[c]{cccc}%
1 & 0 & 0 & 0\\
0 & 1 & 0 & 0\\
0 & 0 & 1 & 0\\
0 & 0 & 0 & -1
\end{array}
\right]  =I_{g_{1}g_{2}} \label{11}%
\end{equation}

which is to label the target state $|g_{1}\rangle|g_{2}\rangle$. Similarily,
target state $|e_{1}\rangle|e_{2}\rangle$ can be labeled by setting
$b=\frac{\pi}{2}$, $h=4m+3$ ($m$ is an interger), i.e., $t_{2}=\frac{\pi
}{4\lambda}$, $\frac{\Omega}{\lambda}=4m+3$, which yields%

\begin{equation}
H^{\otimes2}U_{I}(t_{2})H^{\otimes2}=\left[
\begin{array}
[c]{cccc}%
-1 & 0 & 0 & 0\\
0 & 1 & 0 & 0\\
0 & 0 & 1 & 0\\
0 & 0 & 0 & 1
\end{array}
\right]  =I_{e_{1}e_{2}} \label{12}%
\end{equation}
\newline

As for the target state $|g_{1}\rangle|e_{2}\rangle$ or $|e_{1}\rangle
|g_{2}\rangle$ , they can be achieved by slight modification of the above
operations as follows\cite{13},%

\begin{equation}
\sigma_{x,2}I_{g_{1}g_{2}}\sigma_{x,2}=\left[
\begin{array}
[c]{cccc}%
1 & 0 & 0 & 0\\
0 & 1 & 0 & 0\\
0 & 0 & -1 & 0\\
0 & 0 & 0 & 1
\end{array}
\right]  =I_{g_{1}e_{2}} \label{13}%
\end{equation}

\begin{equation}
\sigma_{x,2}I_{e_{1}e_{2}}\sigma_{x,2}=\left[
\begin{array}
[c]{cccc}%
1 & 0 & 0 & 0\\
0 & -1 & 0 & 0\\
0 & 0 & 1 & 0\\
0 & 0 & 0 & 1
\end{array}
\right]  =I_{e_{1}g_{2}} \label{14}%
\end{equation}

where $\sigma_{x,2}=\left[
\begin{array}
[c]{cc}%
1 & 0\\
0 & 1
\end{array}
\right]  \otimes\left[
\begin{array}
[c]{cc}%
0 & 1\\
1 & 0
\end{array}
\right]  $ is the NOT gate acting on atom 2. Therefore, Given a specific
$\lambda,$ by choosing an appropriate $\Omega$, we can generate all the
two-qubit operations necessary in the two-qubit Grover's algorithm.

To carry out our scheme, we consider a cavity, i.e., a Fabry-Perot resonator
with a single mode and a standing-wave pattern along the cavity axis \cite{5},
as shown in Fig. 1. Two atoms, first simultaneously prepared in box B into
high-lying circular Rydberg state denoted by $|g_{1}\rangle|g_{2}\rangle,$
after $H^{\otimes2}$ operation, are in the initial average state. Then they
undergo the operations in Fig. 1 from the left to the right. For searching
$|g_{1}\rangle|g_{2}\rangle$ or $|e_{1}\rangle|e_{2}\rangle$, our
implementation is straightforward because the atoms interact with the cavity
and the classical field simultaneously. While to search $|e_{1}\rangle
|g_{2}\rangle$ or $|g_{1}\rangle|e_{2}\rangle$, since NOT gates are only
performed on atom 2, we have to employ an inhomogeneous field to distinguish
the two atoms. This can be done by the same trick as in Ref. \cite{5}, i.e.,
introducing an inhomogeneous electric field in the two regions respectively
where $\sigma_{x,2}$\ takes action. Finally the atoms are separately read out
by the state-selective field-ionization detectors D$_{1}$ and D$_{2}.$

Now we briefly discuss the experimental possibility of our proposal.
Considering two Rydberg atoms with principal quantum numbers 50 and 51 and the
radiative time $T_{r}=3\times10^{-2}$ s, we assume that the atom-cavity
coupling stength $g$ is $25\times2\pi$ kHz \cite{7,14}, $\delta=$ $20\times
g,$ and $\Omega\approx20\times\delta.$ Direct calculation shows that the time
for the two $U_{I}$ operations is $4.0\times10^{-4}$ s$.$ As the time for
single-qubit operations is negligible, the implementation time in the cavity
in Fig. 1 is much shorter than the radiative time $T_{r}$. Moreover, the
atomic state evolution is independent of the cavity field state. So our
proposal is realizable with presently available cavity QED techniques.
Furthermore, it should be pointed out that the Rabi frequency $\Omega$ during
the two-qubit gate is about $10\times2\pi$ MHz and should be slightly adjusted
to satisfy the condition $\frac{\Omega}{\lambda}=4m+1$ or $\frac{\Omega
}{\lambda}=4m+3$ mentioned above.

To check the validity of our scheme more strictly, we numerically simulate the
time evolution of the system for finding the target state $|g_{1}\rangle
|g_{2}\rangle$. As shown in Fig. 2 (a), if we assume that the cavity is
initially in a Fock state $|n\rangle,$ the probability of finding the target
state $|g_{1}\rangle|g_{2}\rangle$ slightly decreases with the increase of the
photon number. Even for $n=10,$ however, the fidelity can still be $99.2\%,$
which means the whole process is almost independent of the cavity state. Fig.
2 (b) presents the influence of the imperfect operations on the fidelity. For
simplicity, we assume the initial cavity state to be $|5\rangle$ and the same
imperfection in each pulse. We see from the plot that even for $7\%$ pulse
error, the fidelity is still larger than $90\%.$

In summary, we have proposed a simple scheme for implementing two-qubit Grover
search algorithm in cavity QED. By adding a strong resonant classical field
during the two-qubit operation, we cancelled the phonon-number dependent Stark
shift. Thus our scheme is immune to both cavity decay and the thermal field.
In addition, different from Ref. \cite{5}, in our proposal, two levels
$|g\rangle$ and $|e\rangle$ of each atoms are employed to encode the qubits
(i.e., qubit definitions are the same for both atoms) and all the operations
except the two NOT gates acting on atom 2 are imposed on the two atoms
simultaneouly, which may make the experimental implementation easier.

We are grateful for warmhearted help from Chaohong Li, Yong Li, Qiongtao Xie
and Chenxi Yue. This work is partly supported by National Natural Science
Foundation of China under Grant Nos. 10474118 and 10274093, and partly by the
National Fundamental Research Program of China under Grant No. 2001CB309309.

$.$


\begin{thebibliography}{99}
\bibitem {12}To reach Eq. (9), the matrix elements in Eq. (8) should meet
following condition: $\ \ \ \ \ \ \frac{1}{2}+\frac{1}{2}\cos(bh)e^{-ib}%
=\frac{i}{2}\sin(bh)e^{-ib}=\frac{1}{2}-\frac{1}{2}\cos(bh)e^{-ib}$. It's easy
to have $bh=\frac{\pi}{2}+N\pi\ (N=0,1,2...)$. Therefore, if $N=2m$%
\ with\ $m$\ being\ an integer,$\ $and\ \ $b=\frac{\pi}{2}$, i.e., $\lambda
t=\frac{\pi}{4}$, we have $\Omega t=\frac{\pi}{4}+m\pi$, $\ h=\frac{\Omega
}{\lambda}=4m+1$; If $N=2m+1$\ and\ $b=\frac{3\pi}{2}$, i.e., $\lambda
t=\frac{3\pi}{4}$, we obtain $\Omega t=\frac{\pi}{4}+m\pi+\frac{\pi}{2}$,
$h=\frac{\Omega}{\lambda}=\frac{4m+3}{3}$. Choosing the first solution, we get
Eq. (9) in the shortest time $t_{D}=\frac{\pi}{4\lambda}.$
\end{thebibliography}
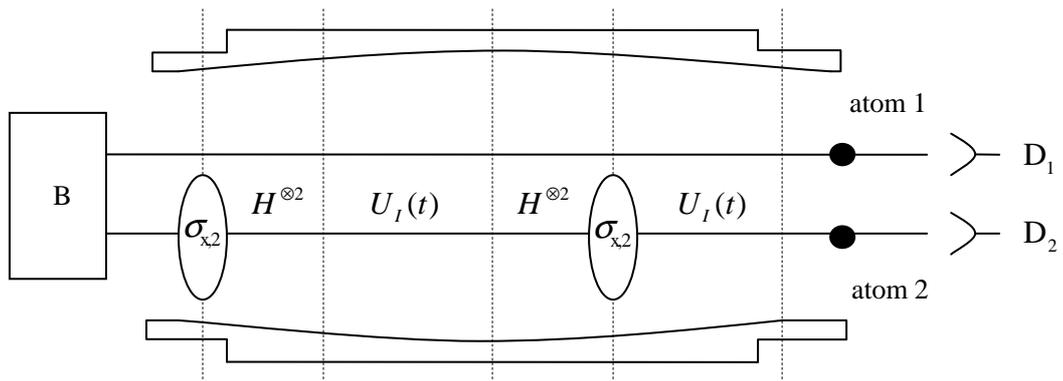
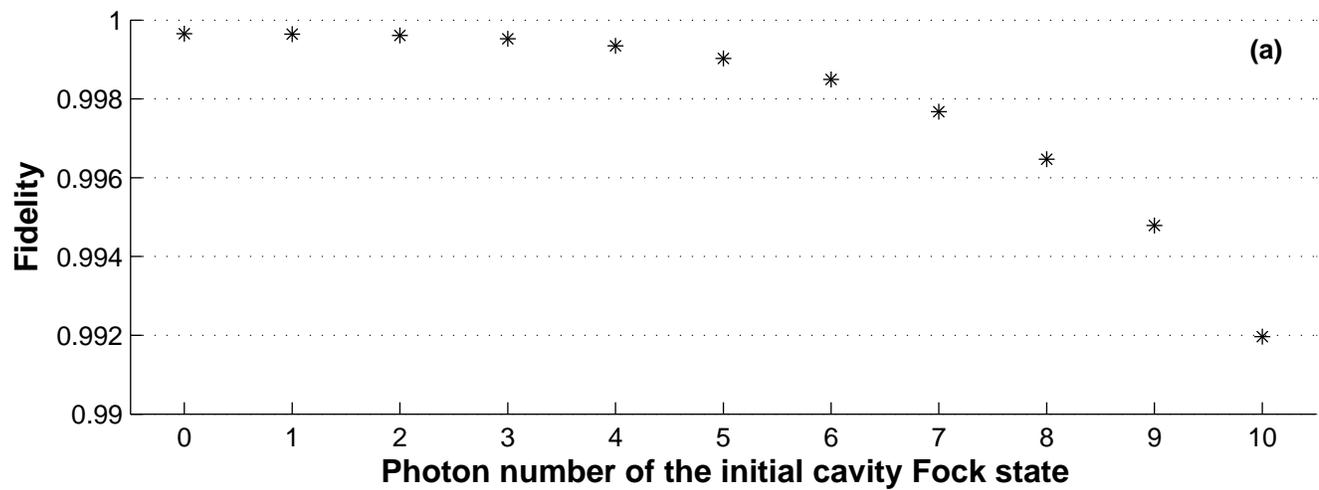
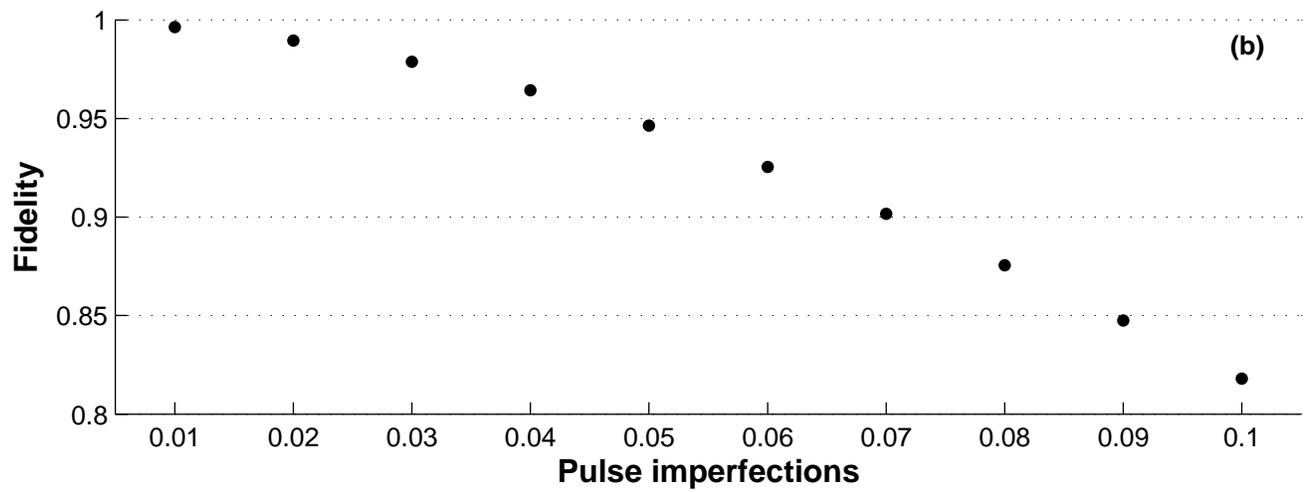
\end{document}